# *Large-Scale Joint Topic, Sentiment & User Preference Analysis for Online Reviews*

Xinli Yu*, Zheng Chen⋆, Wei-Shih Yang*, Xiaohua Hu⋆, Erjia Yan⋆
*Department of Mathematics, Temple University
⋆College of Computing & Informatics, Drexel University

*Abstract*—**This paper presents a non-trivial reconstruction of a previous joint topic-sentiment-preference review model TSPRA with stick-breaking representation under the framework of variational inference (VI) and stochastic variational inference (SVI). TSPRA is a Gibbs Sampling based model that solves topics, word sentiments and user preferences altogether and has been shown to achieve good performance, but for large dataset it can only learn from a relatively small sample. We develop the variational models *vTSPRA* and *svTSPRA* to improve the time use, and our new approach is capable of processing millions of reviews. We rebuild the generative process, improve the rating regression, solve and present the coordinate-ascent updates of variational parameters, and show the time complexity of each iteration is theoretically linear to the corpus size, and the experiments on Amazon datasets show it converges faster than TSPRA and attains better results given the same amount of time. In addition, we tune svTSPRA into an online algorithm *ovTSPRA* that can monitor oscillations of sentiment and preference overtime. Some interesting fluctuations are captured and possible explanations are provided. The results give strong visual evidence that user preference is better treated as an independent factor from sentiment.**

## I. INTRODUCTION

Online *product reviews* have long been a useful subject of research because they provide insight into customer concerns and decision making, and is therefore helping companies to better conduct business activities. One recent challenge is the rapidly increasing amount of reviews, and the trending demand from companies that they want to conduct the analysis on streaming data in a real-time style.

*Topic model* is no doubt a popular method for review analysis [1-5]. When applied to product reviews, a topic is analogous to a *product aspect*, e.g. the battery, or the operating system of a tablet. In topic model, review words are clustered into topics so that we can perform fine-grained analysis like computing topic-level or word-level sentiments, or identifying different levels of customer interests on different product aspects. *Latent Dirichlet Allocation* (LDA) and *Hierarchical Dirichlet Process* (HDP) are two basic frameworks of topic model, where the later can be viewed as an extension of LDA to (countably) infinite many topics [6-8]. HDP has been shown to consistently perform better than LDA [6, 8, 9] "regardless of how many topics LDA uses", but it is of course harder to solve its parameter updates. [3] adapts HDP to model online reviews, because there are many different products with different number of aspects, and it would be tedious to determine them individually. Meanwhile, [3] presents it is to better to model *user preference* (reflecting user interest and concerns) as an independent factor from *word sentiment*, rather than correlate them like [1, 2].

Unfortunately, none of above-mentioned algorithms can process large amount of reviews or steaming data. Many of previous review models are evaluated using *Gibbs sampler* [2, 3, 5], which tends to be more computationally intensive due to the *mixing time* or *burn-in* it takes to reach steady state and produce (asymptotically) exact samples from the target density [10]. Models trained on a small sampled set cannot properly represent a corpus when it is large. They must use meaningless default parameters for users or words not present in the training set and hurt the performance. The sentiment and preference evaluation might also be skewed depending on the corpus structures.

*Variational inference* (VI), or in particular *Mean-Field Approximation* [11, 12] is one way to improve topic model time usage and has been proved effective in speeding up the inference at cost of certain accuracy. It "tends to be faster and easier to scale to large data" [10]. Given a probabilistic model with density $p$, the main idea of VI is to first define a "metric" that measures distance between probability two densities, like the KL-divergence; and then define a set of density functions $Q$ where we search for one $q \in Q$ with minimum distance to $p$: $q^* = \arg\min_{q \in Q} \mathrm{KL}(q|p)$. Thus, VI turns a topic model into a functional optimization problem, and well-studied methods from fields of *stochastic optimization* or *parallel optimization* can come to help scale the algorithm. Stochastic Variational Inference (SVI) [9] [13] scales topic-model VI by randomly sample documents from the corpus and pretends the sample is the whole corpus, and theoretically it is applicable to a corpus of arbitrary size. [9] shows given the same amount of time, SVI learns data better than VI on large datasets.

The ***objective/contribution*** of this paper is to completely rebuild the TSPRA model in [3] to fit big data. The reason that we follow up TSPRA is its several distinctions described in section II.A. We first reconstruct the model using *stick breaking* [6, 8] (rather than CRF) representation of HDP so that it is compatible with VI and SVI in section III.A. We also improve the model design for better parameter interpretation and eliminate unnecessary heuristics. We then solve the analytic solutions in III.B for parameter updates under the VI framework, which is entirely different from what has been given in [3]. The solutions are then further adapted to a SVI algorithm in III.C. In addition, we tune the SVI algorithm to analyze sentiment and preference change overtime, which we believe provide interesting and useful information for real-world business decision making. To our best knowledge, currently there is no previous research of review analysis in this direction.

In this introduction, we have discussed the motivation of this paper and an objective overview. *In Section II* we give a summary of TSPAR and variational inference. *In section III*, we present the model, inference, interpretation and algorithm. *In section IV*, experiments and evaluation results are demoed.

## II. Background

### A. TSPRA

TSPRA [3] proposes a HDP-based framework for joint analysis, called joint Topic-Sentiment-Preference Regression Analysis (TSPRA). It uses the Chinese Restaurant Franchise (CRF) representation of HDP to derive Gibbs sampling formulas. There are several distinctions of this model from other models. Such distinction is also the reason that we follow up this work.

1) It points out the concept of user preference might not be properly reflected in previous models like [1, 2], since they let user preference govern both review ratings and sentiments, it functions more like topic-level sentiment rather than "customer interest" . TSPRA breaks user preferences and sentiments into independent variables, and has identified product aspects with high user preference and low sentiments from Amazon reviews.
2) TSPRA allows a totally automatic word sentiment annotation based on review corpus, and it is the first among topic-based review models to demo word sentiment polarity evaluations, and find the sentiment distribution in review context can differ drastically from general sentiment evaluation like SenticNet [14].
3) It adopts HDP in a review model and remove the need to laboriously find an appropriate topic number for each dataset like in [5].

Due to its inability to process large datasets, in experiments the model is trained with less than 10,000 reviews.

### B. Variational Inference

Mean-field approximation is the most widely used VI method for topic models. It tries to find a best density $q$ with all its marginals being independent to approximate the true complicated distribution defined by the model. In optimization, it uses coordinate ascent to directly jump to the analytically solved optimal choice of each dimension. This method especially fits TSPRA since two additional key variables "user preference" and "word sentiment" are already assumed independent.

- **Theorem 1**. Suppose $\mathbf{x}$ represents all observed data and $\mathbf{Z} = (Z_1, Z_2, \ldots)$ represents all latent variables in a model $p$, then for the $j$th dimension, we have an exact solution
$$q_j(z_j) = \exp\left\{\mathbb{E}_{q_{-z_j}}\left[\log p\left(z_j|\mathbf{z}_{-z_j},\mathbf{x}\right)\right]\right\}$$
where $p\left(z_j|\mathbf{z}_{-z_j},\mathbf{x}\right)$ represents the marginal density of the $j$th dimension given the data and all other dimensions.

In practice $p\left(z_j|\mathbf{z}_{-z_j},\mathbf{x}\right)$ is usually hard to solve, but due to the particular structure of above $q_j$ (the exponential and log function, the linearity of expectation), we can take advantage of $p\left(z_j|\mathbf{z}_{-z_j},\mathbf{x}\right) \propto p\left(z_j,\mathbf{z}_{-z_j},\mathbf{x}\right)$ and then eliminate variables independent from $z_j$. This theorem is repeatedly used without mentioning in section III to derive update formulas.

CRF is used in [3] to construct and interpret HDP. However, variational inference is hard to apply on CRP because some of its complete conditionals do not have an exponential-family form. In this paper, we instead use stick-breaking constructions like [11] [12]; the theoretical

<table>
<tr><td rowspan="6">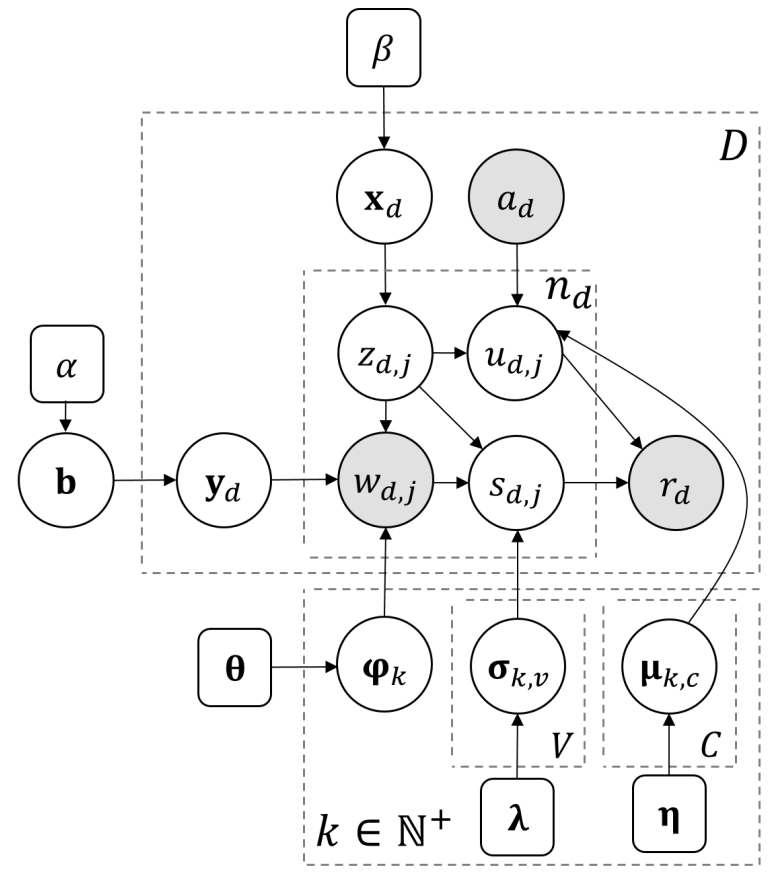

Figure 1 The plate diagram of vTSPRA</td><td>$\alpha$: discount parameter for the corpus level GEM.</td><td>$\beta$: discount parameter for the document level GEM.</td></tr>
<tr><td rowspan="2">$\mathbf{b}, b_k|\tilde{\boldsymbol{\alpha}}_k$: $\mathbf{b}$ are i.i.d drawn from beta$(1,\alpha)$. $\pi(\mathbf{b})$ can then be viewed as an infinite dimensional categorical distribution drawn from GEM$(1,\alpha)$. In the variational model, each of its component $b_k$ obeys beta$(\tilde{\boldsymbol{\alpha}}_k)$.</td><td>$\boldsymbol{\theta}$: Dirichlet parameter of the base distribution from which $V$-dimensional topics are generated.</td></tr>
<tr><td>$\boldsymbol{\varphi}_k|\tilde{\boldsymbol{\theta}}_k$: the $k$th $V$-dimensional topic generated from Dirichlet$(\boldsymbol{\theta})$ . In the variational model, it obeys Dirichlet$(\tilde{\boldsymbol{\theta}}_k)$.</td></tr>
<tr><td>$\boldsymbol{\lambda}$: Dirichlet parameters of the prior distributions of word sentiments.</td><td>$\boldsymbol{\sigma}_{k,v}|\tilde{\boldsymbol{\lambda}}_{k,v}$: the word sentiment distribution for topic $k$ and word $v$ drawn from Dirichlet$(\boldsymbol{\lambda})$; in the variational model it obeys Dirichlet$(\tilde{\boldsymbol{\lambda}}_{k,v})$.</td></tr>
<tr><td>$\boldsymbol{\eta}$: Dirichlet parameters of the prior distributions of user preferences.</td><td>$\boldsymbol{\mu}_{k,c}|\tilde{\boldsymbol{\eta}}_{k,c}$: the user preference distribution for topic $k$ and user $c$ drawn from Dirichlet$(\boldsymbol{\eta})$; in the variational model it obeys Dirichlet$(\tilde{\boldsymbol{\eta}}_{k,c})$.</td></tr>
<tr><td>$\mathbf{y}_d, y_{d,t}|\tilde{\boldsymbol{\xi}}_{d,t}$: $\mathbf{y}_d$ is the document-level infinite topic indices drawn from categorical$(\pi(\mathbf{b}))$ for document $d$. In the variational model, each of its component $b_k$ obeys a categorical distribution parameterized by a probability vector $\tilde{\boldsymbol{\xi}}_{d,t}$.</td><td>$\mathbf{x}_d, x_{d,t}|\tilde{\boldsymbol{\beta}}_{d,t}$: $\mathbf{x}_d$ are i.i.d drawn from beta$(1,\beta)$. $\pi(\mathbf{x}_d)$ can then be viewed as an infinite dimensional categorical distribution drawn from GEM$(1,\beta)$. In the variational model, each of its component $b_k$ obeys beta$(\tilde{\boldsymbol{\beta}}_{d,t})$.</td></tr>
<tr><td>$D$: the number of documents in the corpus; overloaded to represent the corpus as well.</td><td rowspan="2">$w_{d,j}$: the observed $j$th word in document $d$, viewed as being drawn from topic $\boldsymbol{\varphi}_{y_{d,z_{d,j}}}$.</td><td rowspan="2">$z_{d,j}|\tilde{\boldsymbol{\phi}}_{d,j}$: the topic index assignment for word $w_{d,j}$, drawn from categorical$(\pi(\mathbf{x}_d))$. In the variational mode, it obeys categorical$(\tilde{\boldsymbol{\phi}}_{d,j})$.</td></tr>
<tr><td>$n_d$: the number of words in document $d$.</td></tr>
<tr><td>$V$: the vocabulary and its size.</td><td rowspan="3">$s_{d,j}|\tilde{\boldsymbol{\rho}}_{d,j}$: the sentiment assignment for word $w_{d,j}$, drawn from categorical$\left(\boldsymbol{\sigma}_{y_{d,z_{d,j}},w_{d,j}}\right)$ . In the variational mode, it obeys categorical$(\tilde{\boldsymbol{\rho}}_{d,j})$.</td><td rowspan="3">$u_{d,j}|\tilde{\boldsymbol{\upsilon}}_{d,j}$: the user preference assignment for word $w_{d,j}$, drawn from categorical$\left(\boldsymbol{\mu}_{y_{d,z_{d,j}},a_d}\right)$. In the variational mode, it obeys categorical$(\tilde{\boldsymbol{\upsilon}}_{d,j})$.</td></tr>
<tr><td>$C$: the user set and its size.</td></tr>
<tr><td>$\mathcal{S}$: sentiment categories. We use $\{-1,0,1\}$.</td></tr>
<tr><td>$\mathcal{U}$: user preference categories. We use $\{0,1\}$.</td><td>$r_d$: the rating of document $d$, viewed as being drawn from beta$(h(\mathbf{s}_d,\mathbf{u}_d))$.</td><td>$a_d$: the author of document $d$.</td></tr>
</table>

Table 1 Notations of parameters and variables in vTSPRA.

background of such constructions is explained in [8] with more detail.

## III. Model & Inference

### A. Model Formulation

We first build up the generative process of model *vTSPRA*, the variational TSPRA, using stick breaking. The plate diagram and notation are summarized in Table 1. Following convention, bold letters represent vector or matrix. Some letters like $D, V, C$, are overloaded to both represent the corpus, vocabulary, user set and the size of them.

***Review generation***. The stick-breaking construction needs a $V$-dimensional Dirichlet distribution $\text{Dirichlet}(\boldsymbol{\theta})$ as the so-called base distribution from which categorical distributions representing topics are generated, or simply the prior distribution of topics. Then, it assumes every word $w_{d,j}, j = 1, \dots, n_d$ in every review document $d = 1, \dots, D$ are generated by a process analogous to an infinite dimensional Latent Dirichlet Allocation.

- Draw countably infinite many topics
$$\boldsymbol{\varphi}_k \sim \text{Dirichlet}(\boldsymbol{\theta}), k \in \mathbb{N}^+$$
- Draw corpus-level stick breaking $b_k \sim \text{beta}(1, \alpha), k \in \mathbb{N}^+$.
- For each document $d$,
  - Draw document-level topic indices
$$y_{d,t} \sim \text{categorical}(\pi(\mathbf{b})), t \in \mathbb{N}^+$$
  - Draw document-level stick-breaking
$$x_{d,t} \sim \text{beta}(1, \beta), t \in \mathbb{N}^+$$
  - For each word index $i$ of $d$,
    - Draw topic assignment $z_{d,i} \sim \text{categorical}(\pi(\mathbf{x}_d))$
    - Draw word $w_{d,i} \sim \text{categorical}\left(\boldsymbol{\varphi}_{y_{d,z_{di}}}\right)$

***Rating regression***. For mathematical convenience, we first normalize all ratings to range $[0,1]$. As a basic example, if the minimum rating is $r_l$, the maximum rating is $r_h$, and the original rating is $r^*$, then a straightforwardly normalized rating would be $r = \frac{r^* - r_l}{r_h - r_l}$. Using this formula, a 3/5 Amazon review rating is 0.5. In practice, we should actually normalize it to $[\epsilon, 1 - \epsilon]$ where $\epsilon$ is a very small quantity like $\epsilon = 10^{-300}$ whose logarithm is not negative infinity on the computation platform; e.g. an Amazon review of rating 1 is mapped to $\epsilon$ rather than zero.

We then denote the set of all possible user preferences as $\mathcal{U}$, and the set of all possible word sentiments as $\mathcal{S}$, where we restrict $\mathcal{U} \subseteq [0,1]$ and $\mathcal{S} \subseteq [-1,1]$. In this paper, we let $\mathcal{U} = \{0.5, 1\}$, representing weak/strong user preference, and $\mathcal{S} = \{-1, 0, 1\}$, representing negative/neutral/positive sentiment. Of course, more levels can be added if desired. We assume the rating $r_d$ of each review $d$ is regressed by the word sentiments $\mathbf{s}_d = (s_{d,j})$ and the user preferences $\mathbf{u}_d = (u_{d,j})$ in $d$ by some customized function $h(\mathbf{s}_d, \mathbf{u}_d): \mathcal{S} \times \mathcal{U} \to (0, +\infty)^2$. The entire rating generation process is designed as the following,

- For each topic $k$,
  - For each unique word $v$ in the vocabulary $V$, draw a $|\mathcal{S}|$-dimensional probability vector $\boldsymbol{\sigma}_{k,v} \sim \text{Dirichlet}(\boldsymbol{\lambda})$. In this paper $|\mathcal{S}| = 3$.
  - For each unique user $c$ in the user set $C$, draw a $|\mathcal{U}|$-dimensional probability vector $\boldsymbol{\mu}_{k,c} \sim \text{Dirichlet}(\boldsymbol{\eta})$. In this paper $|\mathcal{U}| = 2$.
- For each document $d$,
  - For each word index $j$ of $d$,
    - Draw word sentiment $s_{d,j} \sim \text{categorical}\left(\boldsymbol{\sigma}_{y_{d,z_{d,j}}, w_{d,j}}\right)$;
    - Draw user preference $u_{d,j} \sim \text{categorical}\left(\boldsymbol{\mu}_{y_{d,z_{d,j}}, c_d}\right)$.
  - Draw $r_d \sim \text{beta}(h_1, h_2)$ where $h_1, h_2$ are two dimensions of $h(\mathbf{s}_d, \mathbf{u}_d)$. In this paper $h$ is given by (2).

We will see later there is a subtle reason for interpretation of variational parameters of preferences and sentiments. Also, this regression design keeps the independence between $u_{d,j}$ and $s_{d,j}$, which can be verified by D-separation theorem. Besides that, it considers computation complexity; we show later our variational inference of this model is linear with the corpus size.

For $h$, we view the review rating as a noisy average of word rating $r_{d,j}$, and we employ a heuristic $r_{dj} = u_{d,j} s_{d,j}$. By this perspective and beta distribution properties, $\mathbb{E}[r_d] = \frac{h_1}{h_1 + h_2}$ can be estimated by mean word ratings $\text{mean}(r_{d,j})$, and since $\text{var}(r_d) = \frac{h_1 h_2}{(h_1 + h_2 + 1)(h_1 + h_2)^2}$, then $h_1 + h_2$ can be estimated by

$$h_1 + h_2 \approx \frac{\text{mean}(r_{d,j}) \times \left(1 - \text{mean}(r_{d,j})\right)}{\text{var}(r_{d,j})} - 1 \qquad (1)$$

from which we conclude

$$h(\mathbf{s}_d, \mathbf{u}_d) = \left( \left(\text{mean}(r_{dj}), 1 - \text{mean}(r_{dj})\right) \times \left( \frac{\text{mean}(r_{d,j}) \times \left(1 - \text{mean}(r_{d,j})\right)}{\text{var}(r_{d,j})} - 1 \right) \right) \qquad (2)$$

A major difference between this paper's rating regression and [3] is that review rating is now drawn from beta distribution instead of Gaussian distribution. This is a more reasonable choice, as beta distribution can be restricted to a range, not to extend to infinity, and is more flexible in its shape with also two parameters. For example, a uniform or near-uniform distribution occurs quite often in our case when there is no strong evidence neither for high rating nor for low rating. Gaussian distribution is awkward in mimicking uniform distribution, where one can imagine that it must place much of its density outside the rating range; in contrast, a beta distribution with both of its parameters being 1 is straightly uniform.

***As a summary*** of both parts, in the entire model, the corpus-level latent variables include the corpus-level stick breaking $\mathbf{b}$, the topics $\{\boldsymbol{\varphi}_k\}$, the topic-word sentiments $\{\boldsymbol{\sigma}_{k,v}\}$ and the user-topic preferences $\{\boldsymbol{\mu}_{k,c}\}$; the document-level latent variables include document-level topic indices $\{y_{d,k}\}$, document-level stick-breaking $\{\mathbf{x}_d\}$; local latent variables include topic assignment $\{z_{d,j}\}$, user preference assignment $\mathbf{u}_d = (u_{d,j})$, and word sentiment assignment $\mathbf{s}_d = (s_{d,j})$. The observables are the review document texts $\mathbf{w}_1, \dots, \mathbf{w}_D$, ratings $r_1, \dots, r_D$ and their authors $a_1, \dots, a_D$. The density for all latent variables and observables given other parameters can be written as

$$p = p_c \times \prod_{d=1}^{D} \left( p_d \times \left( \prod_{j=1}^{n_d} p_{d,j} \right) \right) \tag{3}$$

where

$$\begin{aligned} p_c &= p(\mathbf{b}|\alpha) \\ &\times \prod_{k=1}^{\infty} \left( p(\boldsymbol{\varphi}_k|\boldsymbol{\theta}) \times \prod_{c=1}^{C} p(\boldsymbol{\mu}_{k,c}|\boldsymbol{\eta}) \times \prod_{v=1}^{V} p(\boldsymbol{\sigma}_{k,v}|\boldsymbol{\lambda}) \right) \end{aligned} \tag{4}$$

$$p_d = p(\mathbf{x}_d|\beta) \times \left( \prod_{t=1}^{\infty} p\left( y_{d,t}|\pi(\mathbf{b}) \right) \right) \times p(r_d|\mathbf{u}_d, \mathbf{s}_d) \tag{5}$$

$$\begin{aligned} p_{d,j} &= p\left( z_{d,j}|\pi(\mathbf{x}_d) \right) \times p\left( w_{d,j}|\boldsymbol{\varphi}_{y_{d,z_{d,j}}} \right) \\ &\times p\left( u_{d,j}|\boldsymbol{\mu}_{y_{d,z_{d,j}}, a_d} \right) \times p\left( s_{d,j}|\boldsymbol{\sigma}_{y_{d,z_{d,j}}, w_{d,j}} \right) \end{aligned} \tag{6}$$

## *B. Variational Inference*

The density of latent variables given observed data $\mathbf{w}, \mathbf{r}, \mathbf{a}$ in our case is $p(\mathbf{b}, \mathbf{x}, \boldsymbol{\varphi}, \boldsymbol{\mu}, \boldsymbol{\sigma}, \mathbf{y}, \mathbf{z}, \mathbf{u}, \mathbf{s}|\mathbf{w}, \mathbf{r}, \mathbf{a})$.The mean-field approximation restricts densities in $Q$ to have independent marginals. In particular, $q \in Q$ can be written as the following with $\tilde{\boldsymbol{\alpha}}, \tilde{\boldsymbol{\beta}}, \tilde{\boldsymbol{\theta}}, \tilde{\boldsymbol{\eta}}, \tilde{\boldsymbol{\lambda}}, \tilde{\boldsymbol{\xi}}, \tilde{\boldsymbol{\phi}}, \tilde{\boldsymbol{\upsilon}}, \tilde{\boldsymbol{\rho}}$ as corresponding variational parameters.

$$q = q_c \times \prod_{d=1}^{D} \left( q_d \times \left( \prod_{j=1}^{n_d} q_{d,j} \right) \right) \tag{7}$$

where

$$q_c = \prod_{k=1}^{K} q(\boldsymbol{\varphi}_k|\tilde{\boldsymbol{\theta}}_k) q(b_k|\tilde{\boldsymbol{\alpha}}_k) \times \prod_{c=1}^{C} q(\boldsymbol{\mu}_{k,c}|\tilde{\boldsymbol{\eta}}_{k,c}) \prod_{v=1}^{V} q(\boldsymbol{\sigma}_{k,v}|\tilde{\boldsymbol{\lambda}}_{k,v}) \tag{8}$$

$$q_d = \prod_{t=1}^{T} q(x_{d,t}|\tilde{\boldsymbol{\beta}}_{d,t})\, q(y_{d,t}|\tilde{\boldsymbol{\xi}}_{d,t}) \tag{9}$$

$$q_{d,j} = q(z_{d,j}|\tilde{\boldsymbol{\phi}}_{d,j}) q(u_{d,j}|\tilde{\boldsymbol{\upsilon}}_{d,j}) \times q(s_{d,j}|\tilde{\boldsymbol{\rho}}_{d,j}) \tag{10}$$

Here we truncate corpus-level number of topics by $K$, and document-level number of topics by $T$. Note topic number truncation is done on the variational density [10], not the original model; they can be set to a reasonable upper bound of number of topics. [8] includes formulas for mean and variance of HDP; based on that and experiment results of [3] we would recommend $K = 100$ and $T = 10$ for Amazon reviews. In addition, we require each marginal of $q$ to have the same support of its counterpart of $p$. This implies $q(u_{d,j}|\tilde{\boldsymbol{\upsilon}}_{d,j})$ and $q(s_{d,j}|\tilde{\boldsymbol{\rho}}_{d,j})$ are categorical distribution over the same $\mathcal{U}$ and $\mathcal{S}$ respectively. Together with truncation, it also implies $q(y_{d,t}|\tilde{\boldsymbol{\xi}}_{d,t})$ is a categorical over $1, \dots, K$, and $q(z_{d,j}|\tilde{\boldsymbol{\phi}}_{d,j})$ is a categorical over $1, \dots, T$. We note the marginal independence, the truncation and the same-support requirement are the only necessary restrictions on $Q$. There is no need to pre-assume like $q(b_k|\tilde{\boldsymbol{\alpha}}_k)$ is beta or $q(\boldsymbol{\sigma}_{k,v}|\tilde{\boldsymbol{\lambda}}_{k,v})$ is Dirichlet like in [9].

Marginals of $q$ are represented by corresponding variational parameters. We use coordinate ascent of each marginal's variational parameters as in [10] [11] to find a local optimal for $q$. The mathematical trick of using identity random variables to separate a random vector and its random indices is repeatedly applied in order to compute expectation.

***Update of*** $\tilde{\boldsymbol{\theta}}$. From (4) and (6) we have the complete conditional

$$\begin{aligned} & p(\boldsymbol{\varphi}_k|\sim) \\ & \propto p(\boldsymbol{\varphi}_k|\boldsymbol{\theta}) \prod_{d=1}^{D} \prod_{j=1}^{n_d} p\left( w_{d,j}|\boldsymbol{\varphi}_{y_{d,z_{d,j}}} \right)^{\mathbb{I}(k = y_{d,z_{d,j}})} \\ & = p(\boldsymbol{\varphi}_k|\boldsymbol{\theta}) \prod_{d=1}^{D} \prod_{t=1}^{\infty} \left( \prod_{j=1}^{n_d} \varphi_{k,w_{d,j}}{}^{\mathbb{I}(t = z_{d,j})} \right)^{\mathbb{I}(k = y_{d,t})} \end{aligned} \tag{11}$$

Then by Theorem 1 we have

$$\begin{aligned} q(\boldsymbol{\varphi}_k) &\propto \exp\left\{ \mathbb{E}_{q_{-\boldsymbol{\varphi}_k}} \left[ \log p(\boldsymbol{\varphi}_k|\boldsymbol{\theta}) \prod_{d=1}^{D} \prod_{t=1}^{\infty} \left( \prod_{j=1}^{n_d} \varphi_{k,w_{d,j}}{}^{\mathbb{I}(t = z_{d,j})} \right)^{\mathbb{I}(k = y_{d,t})} \right] \right\} \\ &= \exp\left\{ \log p(\boldsymbol{\varphi}_k|\boldsymbol{\theta}) + \mathbb{E}_{q_{-\boldsymbol{\varphi}_k}} \left[ \sum_{d=1}^{D} \sum_{t=1}^{\infty} \mathbb{I}(k = y_{d,t}) \sum_{j=1}^{n_d} \mathbb{I}(t = z_{d,j}) \log \varphi_{k,w_{d,j}} \right] \right\} \\ &= \exp\left\{ \log p(\boldsymbol{\varphi}_k|\boldsymbol{\theta}) \right. \\ &\quad \left. + \sum_{d=1}^{D} \sum_{t=1}^{T} \mathbb{E}_{q_{-\boldsymbol{\varphi}_k}}[\mathbb{I}(k = y_{d,t})] \sum_{j=1}^{n_d} \mathbb{E}_{q_{-\boldsymbol{\varphi}_k}}[\mathbb{I}(t = z_{d,j})] \mathbb{E}_{q_{-\boldsymbol{\varphi}_k}}\left[ \log \varphi_{k,w_{d,j}} \right] \right\} \\ &= \exp\left\{ \log p(\boldsymbol{\varphi}_k|\boldsymbol{\theta}) + \sum_{d=1}^{D} \sum_{t=1}^{T} \tilde{\boldsymbol{\xi}}_{d,t}(k) \sum_{j=1}^{n_d} \tilde{\boldsymbol{\phi}}_{d,j}(t) \log \varphi_{k,w_{d,j}} \right\} \end{aligned} \tag{12}$$

where the third identity is by independence of $\mathbf{z}, \mathbf{y}, \boldsymbol{\varphi}$ in the variational model, and the last identity is by the following fact,

- **Theorem 2**. The expectation of an identity random variable with respect a probability measure equals its probability of that identity under that measure.

For example, above theorem implies $\mathbb{E}_{q_{-\boldsymbol{\varphi}_k}}[\mathbb{I}(k = y_{d,t})] = q_{-\boldsymbol{\varphi}_k}\{y_{d,t} = k\}$, and we have shown $y_{d,t}$ is categorical parameterized by a probability vector $\tilde{\boldsymbol{\xi}}_{d,t}$ under $q_{-\boldsymbol{\varphi}_k}$, then clearly $\mathbb{E}_{q_{-\boldsymbol{\varphi}_k}}[\mathbb{I}(k = y_{d,t})] = \tilde{\boldsymbol{\xi}}_{d,t}(k)$. For the same reason $\mathbb{E}_{q_{-\boldsymbol{\varphi}_k}}\left[ \log \varphi_{k,w_{d,j}} \right] = \tilde{\boldsymbol{\phi}}_{d,j}(t)$. This fact will be implicitly and repeatedly applied in future inference.

Another well-known fact is that $\boldsymbol{\varphi}_k \sim \text{Dirichlet}(\boldsymbol{\theta})$ is an exponential family with $\boldsymbol{\theta}$ itself as the natural parameter and $\log \boldsymbol{\varphi}_k$ as the sufficient statistics. It is not hard to find from the final form of (12) that $q(\boldsymbol{\varphi}_k)$ is also written in exponential-family form with $\log \boldsymbol{\varphi}_k$ as its sufficient statistics. As a result, $q(\boldsymbol{\varphi}_k)$ is the density of a Dirichlet distribution, and some simple arithmetic will reveal its natural parameter, which is also the variational parameter $\tilde{\boldsymbol{\theta}}_k$, is

$$\tilde{\boldsymbol{\theta}}_k = \boldsymbol{\theta} + \sum_{d=1}^{D}\sum_{t=1}^{T} \tilde{\boldsymbol{\xi}}_{d,t}(k) \sum_{j=1}^{n_d} \tilde{\boldsymbol{\phi}}_{d,j}(t)\mathbf{e}_{w_{d,j}} \tag{13}$$

where $\mathbf{e}_{w_{d,j}}$ is a $V$-dimensional vector s.t. all its components are zero except for the $w_{d,j}$th component is 1. The time complexity for (12) is $O\left(K(T(\sum_{d=1}^{D} n_d) + V)\right) = O(KTN)$ given that $N = \sum_{d=1}^{D} n_d$ and $V \ll N$.

***Updates of*** $\tilde{\boldsymbol{\lambda}}$ ***and*** $\tilde{\boldsymbol{\eta}}$ use the same technique as above. $\boldsymbol{\sigma}_{k,v}$ and $\boldsymbol{\mu}_{k,c}$ still obeys Dirichlet distribution in the variational model, and the results for natural parameter updates are given below:

$$\tilde{\boldsymbol{\lambda}}_{k,v} = \boldsymbol{\lambda} + \sum_{d=1}^{D}\sum_{t=1}^{T} \tilde{\boldsymbol{\xi}}_{d,t}(k) \sum_{j=1}^{n_d} \mathbb{I}(v = w_{d,j})\tilde{\boldsymbol{\phi}}_{d,j}(t)\tilde{\boldsymbol{\rho}}_{d,j} \tag{14}$$

$$\tilde{\boldsymbol{\eta}}_{k,c} = \boldsymbol{\eta} + \sum_{d=1}^{D}\sum_{t=1}^{T} \tilde{\boldsymbol{\xi}}_{d,t}(k) \sum_{j=1}^{n_d} \mathbb{I}(c = a_d)\tilde{\boldsymbol{\phi}}_{d,j}(t)\tilde{\boldsymbol{\upsilon}}_{d,j} \tag{15}$$

In implementation, it is not hard to update both at $O(KTN)$ complexity not dependent on $V$ or $C$, since we can update corresponding entries of $\tilde{\boldsymbol{\lambda}}$ and $\tilde{\boldsymbol{\eta}}$ for each review and each word. This is ***important*** since $V, C$ grows with $N$ at a power of 0.5 to 0.7 on the Amazon dataset, effectively making the complexity non-linear if it is dependent on them. Moreover, we have an *interpretation and use* for $\boldsymbol{\sigma}_{k,v} \sim \text{Dirichlet}(\tilde{\boldsymbol{\lambda}}_{k,v})$ and $\boldsymbol{\mu}_{k,c} \sim \text{Beta}(\tilde{\boldsymbol{\eta}}_{k,c})$ in the variational model.

- Since $\mathbb{E}_q[\boldsymbol{\sigma}_{k,v}] = \frac{\tilde{\boldsymbol{\lambda}}_{k,v}}{\|\tilde{\boldsymbol{\lambda}}_{k,v}\|_1}, \mathbb{E}_q[\boldsymbol{\mu}_{k,c}] = \frac{\tilde{\boldsymbol{\eta}}_{k,c}}{\|\tilde{\boldsymbol{\eta}}_{k,c}\|_1}$, then $\frac{\tilde{\boldsymbol{\lambda}}_{k,v}}{\|\tilde{\boldsymbol{\lambda}}_{k,v}\|_1}$ can be naturally interpreted as the mean probability weights whether word $v$ is negative, neutral or positive under topic $k$, and likewise $\frac{\tilde{\boldsymbol{\eta}}_{k,c}}{\|\tilde{\boldsymbol{\eta}}_{k,c}\|_1}$ can be interpreted as the mean probability weights whether customer $c$ has weak or strong preference to topic $k$. We denote

$$\hat{\boldsymbol{\sigma}}_{k,v} = \frac{\tilde{\boldsymbol{\lambda}}_{k,v}}{\|\tilde{\boldsymbol{\lambda}}_{k,v}\|_1}, \hat{\boldsymbol{\mu}}_{k,c} = \frac{\tilde{\boldsymbol{\eta}}_{k,c}}{\|\tilde{\boldsymbol{\eta}}_{k,c}\|_1} \tag{16}$$

  as our estimates of sentiment of $v$ and preference of user $c$ w.r.t. topic $k$. The estimate of overall sentiment of word $v$ or topic $k$ can thus be

$$\hat{\sigma}_v = \frac{\mathscr{s}^{\mathrm{T}}(\sum_{k=1}^{K} \tilde{\boldsymbol{\lambda}}_{k,v})}{\sum_{k=1}^{K}\|\tilde{\boldsymbol{\lambda}}_{k,v}\|_1}, \hat{\sigma}_k = \frac{\mathscr{s}^{\mathrm{T}}\left(\sum_{v\in V} \tilde{\boldsymbol{\lambda}}_{k,v}\right)}{\sum_{v\in V}\|\tilde{\boldsymbol{\lambda}}_{k,v}\|_1} \tag{17}$$

  where $\mathscr{s}$ is a vector with components being the corresponding sentiment level, and is $\mathscr{s} = (-1,0,1)$ in this paper.
- The 1-norm of the parameter vector of Dirichlet distribution is called its concentration. A higher concentration implies the distribution places more of its density in the local neighborhood around the mean. In our case, a higher $\|\tilde{\boldsymbol{\lambda}}_{k,v}\|_1, \|\tilde{\boldsymbol{\eta}}_{k,c}\|_1$ implies the variational model $q$ places more density near the estimates $\hat{\sigma}_{k,v}$ and $\hat{\mu}_{k,c}$. In plain words, we can therefore trust more in the estimates $\hat{\sigma}_{k,v}$ and $\hat{\mu}_{k,c}$.

In comparison, [3] estimates the word sentiments and user preferences using a heuristic formula based on the number of occurrences of positive/negative sentiments, and the number of occurrences of positive/negative user preferences. Such formulas are less mathematically & computationally graceful in the sense that they are based on hunch with little theoretical support, and they have to do the counts in another iteration after all inference of latent variables are done.

***Update of*** $\tilde{\boldsymbol{\alpha}}$, the variational parameters for corpus-level stick breaking $\mathbf{b}$. From (4) and (5) we can find the complete conditional as we did for (11),

$$p(b_k) \propto p(b_k|\alpha) \prod_{d=1}^{D}\prod_{t=1}^{\infty} p\left(y_{d,t}|\pi(\mathbf{b})\right) \tag{18}$$

and then by Theorem 1 have

$$q(b_k) \propto \exp\left\{\log p(b_k|\alpha) + \sum_{d=1}^{D}\sum_{t=1}^{\infty} \mathbb{E}_{q_{-b_k}}\left[\log p\left(y_{d,t}|\pi(\mathbf{b})\right)\right]\right\} \tag{19}$$

First, by the density of beta distribution,

$$\begin{aligned}\log p(b_k|\alpha) &= \alpha \log(1-b_k) - \log\frac{\Gamma(\alpha)}{\Gamma(1+\alpha)} - \log(1-b_k) \\ &\propto (\alpha-1)\log(1-b_k)\end{aligned} \tag{20}$$

Secondly, by the definition of GEM,

$$\begin{aligned}p\left(y_{d,t}\middle|\pi(\mathbf{b})\right) &= (1-b_1)\ldots(1-b_{y_{d,t}-1})\times b_{y_{d,t}} \\ &= \left(\prod_{\mathscr{k}=1}^{\infty}(1-b_{\mathscr{k}})^{\mathbb{I}(y_{d,t}>\mathscr{k})}\right)\times\left(\prod_{\mathscr{k}=1}^{\infty} b_{\mathscr{k}}^{\mathbb{I}(y_{d,t}=\mathscr{k})}\right)\end{aligned} \tag{21}$$

which leads to

$$\begin{aligned}&\mathbb{E}_{q_{-b_k}}[\log p(y_{d,t}|\mathbf{b})] \\ &= \mathbb{E}_{q_{-b_k}}\left[\log \prod_{\mathscr{k}=1}^{\infty}(1-b_{\mathscr{k}})^{\mathbb{I}(y_{d,t}>\mathscr{k})} b_{\mathscr{k}}^{\mathbb{I}(y_{d,t}=\mathscr{k})}\right] \\ &= \sum_{\mathscr{k}=1}^{K} \mathbb{E}_{q_{-b_k}}\left[\mathbb{I}(y_{d,t}=\mathscr{k})\log b_{\mathscr{k}}\right] + \mathbb{E}_{q_{-b_k}}\left[\mathbb{I}(y_{d,t}>\mathscr{k})\log(1-b_{\mathscr{k}})\right] \\ &= \sum_{\mathscr{k}=1}^{K} q(y_{d,t}=\mathscr{k})\mathbb{E}_{q_{-b_k}}[\log b_{\mathscr{k}}] + q(y_{d,t}>\mathscr{k})\mathbb{E}_{q_{-b_k}}[\log(1-b_{\mathscr{k}})]\end{aligned} \tag{22}$$

where truncation is applied in the second identity of (22) to make it a finite sum, and the last identity of (22) is due to law of total expectation. Plug (20) and (22) back to (19) and note $q(y_{d,t} = \mathscr{k}) = \tilde{\boldsymbol{\xi}}_{d,t}(\mathscr{k})$ for any $\mathscr{k} = 1, \ldots, K$ since $y_{d,t}$ is categorical in $q$, then

$$\begin{aligned}q(b_k) &\propto \exp\left\{\log p(b_k) + \sum_{d=1}^{D}\sum_{t=1}^{T}\sum_{\mathscr{k}=1}^{K} q(y_{d,t}=\mathscr{k})\mathbb{E}_{q_{-b_k}}[\log b_{\mathscr{k}}] + q(y_{d,t}>\mathscr{k})\mathbb{E}_{q_{-b_k}}[\log(1-b_{\mathscr{k}})]\right\} \\ &\propto \exp\left\{\log p(b_k) + \sum_{d=1}^{D}\sum_{t=1}^{T} q(y_{d,t}=k)\mathbb{E}_{q_{-b_k}}[\log b_k] + q(y_{d,t}>k)\mathbb{E}_{q_{-b_k}}[\log(1-b_k)]\right\} \\ &\propto \exp\left\{\left(\sum_{d=1}^{D}\sum_{t=1}^{T}\tilde{\boldsymbol{\xi}}_{d,t}(k), \alpha - 1 + \sum_{d=1}^{D}\sum_{t=1}^{T}\sum_{\mathscr{k}=k+1}^{K}\tilde{\boldsymbol{\xi}}_{d,t}(\mathscr{k})\right)\begin{pmatrix}\log b_k \\ \log(1-b_k)\end{pmatrix}\right\}\end{aligned} \tag{23}$$

We see $b_k$ still obeys beta distribution in $q$ by comparing (23) with the exponential-family form of beta distribution density, and it immediately follows that

$$\tilde{\boldsymbol{\alpha}}_k = \begin{pmatrix} \sum_{d=1}^{D}\sum_{t=1}^{T} \tilde{\boldsymbol{\xi}}_{d,t}(k) \\ \alpha - 1 + \sum_{d=1}^{D}\sum_{t=1}^{T}\sum_{\ell=k+1}^{K} \tilde{\boldsymbol{\xi}}_{d,t}(\ell) \end{pmatrix} \quad (24)$$

The time complexity for update of $\tilde{\boldsymbol{\alpha}}$ is $O(DTK)$ since the last summation $\sum_{\ell=k+1}^{K}(\cdot)$ in $\tilde{\boldsymbol{\alpha}}_k(2)$ can be computed in a cumulative manner when updating $\tilde{\boldsymbol{\alpha}}_k(1)$.

***Update of*** $\tilde{\boldsymbol{\beta}}$, the variational parameters for document-level stick breakings $\mathbf{x}_d$, can be found in the same way as the following with time complexity $O(TN)$.

$$\tilde{\boldsymbol{\beta}}_{d,t} = \begin{pmatrix} \sum_{j=1}^{n_d} \tilde{\boldsymbol{\phi}}_{d,j}(t) \\ \beta - 1 + \sum_{j=1}^{n_d}\sum_{\ell=t+1}^{T} \tilde{\boldsymbol{\phi}}_{d,j}(\ell) \end{pmatrix} \quad (25)$$

***Update of*** $\tilde{\boldsymbol{\xi}}$, the variational parameters for topic indices per document $\mathbf{y}_d$. We first have

$$\begin{aligned} \tilde{\boldsymbol{\xi}}_{d,t} = q(y_{d,t}) \propto \exp\Bigg\{ & \mathbb{E}_{q_{-y_{d,t}}}[\log p(y_{d,t}|\mathbf{b})] \\ & + \mathbb{E}_{q_{-y_{d,t}}}\left[\log\prod_{j=1}^{n_d}\left(p\left(w_{d,j}|\boldsymbol{\varphi}_{y_{d,z_{d,j}}}\right)\right)^{\mathbb{I}(t=z_{d,j})}\right] \\ & + \mathbb{E}_{q_{-y_{d,t}}}\left[\log\prod_{j=1}^{n_d}\left(p\left(u_{d,j}|\boldsymbol{\mu}_{y_{d,z_{d,j}},a_d}\right)\right)^{\mathbb{I}(t=z_{d,j})}\right] \\ & + \mathbb{E}_{q_{-y_{d,t}}}\left[\log\prod_{j=1}^{n_d}\left(p\left(s_{d,j}|\boldsymbol{\sigma}_{y_{d,z_{d,j}},w_{d,j}}\right)\right)^{\mathbb{I}(t=z_{d,j})}\right]\Bigg\} \end{aligned} \quad (26)$$

For the first expectation, we have

$$\begin{aligned} & \mathbb{E}_{q_{-y_{d,t}}}[\log p(y_{d,t}|\mathbf{b})] \\ & = \sum_{\ell=1}^{\infty} \mathbb{I}(y_{d,t} = \ell)\mathbb{E}_{q(\mathbf{b}(\ell))}[\log \mathbf{b}(\ell)] \\ & \qquad + \mathbb{I}(y_{d,t} > \ell)\mathbb{E}_{q(\mathbf{b}(\ell))}[\log(1-\mathbf{b}(\ell))] \\ & = \mathbb{E}_{q(\mathbf{b}(y_{d,t}))}[\log \mathbf{b}(y_{d,t})] + \sum_{\ell=1}^{y_{d,t}-1} \mathbb{E}_{q(\mathbf{b}(\ell))}[\log(1-\mathbf{b}(\ell))] \\ & = \Psi\left(\tilde{\boldsymbol{\alpha}}_{y_{d,t}}(1)\right) - \Psi\left(\left\|\tilde{\boldsymbol{\alpha}}_{y_{d,t}}\right\|_1\right) + \sum_{\ell=1}^{y_{d,t}-1} \Psi(\tilde{\boldsymbol{\alpha}}_\ell(2)) - \Psi(\|\tilde{\boldsymbol{\alpha}}_\ell\|_1) \end{aligned} \quad (27)$$

where the last identity is using the result of (23) that $\mathrm{b}(y_{d,t})$ obeys beta distribution and the fact that the expectation of a logged beta/Dirichlet random variable can be written in terms of Digamma functions [9, 15]. For the second expectation,

$$\begin{aligned} & \mathbb{E}_{q_{-y_{d,t}}}\left[\log\prod_{j=1}^{n_d}\left(p\left(w_{d,j}|\boldsymbol{\varphi}_{y_{d,z_{d,j}}}\right)\right)^{\mathbb{I}(t=z_{d,j})}\right] = \mathbb{E}_{q_{-y_{d,t}}}\left[\sum_{j=1}^{n_d}\mathbb{I}(t=z_{d,j})\log\varphi_{y_{d,t},w_{d,j}}\right] \\ & = \sum_{j=1}^{n_d}\mathbb{E}_{q_{-y_{d,t}}}\left[\mathbb{I}(t=z_{d,j})\log\prod_{k=1}^{\infty}\left(\varphi_{k,w_{d,j}}\right)^{\mathbb{I}(k=y_{d,t})}\right] \\ & = \sum_{j=1}^{n_d}\sum_{k=1}^{K}\mathbb{I}(k=y_{d,t})\mathbb{E}_{q_{-y_{d,t}}}[\mathbb{I}(t=z_{d,j})]\mathbb{E}_{q_{-y_{d,t}}}\left[\log\left(\varphi_{k,w_{d,j}}\right)\right] \\ & = \sum_{j=1}^{n_d}\tilde{\boldsymbol{\phi}}_{d,j}(t)\left(\Psi\left(\tilde{\boldsymbol{\theta}}_{y_{d,t}}(w_{d,j})\right) - \Psi\left(\left\|\tilde{\boldsymbol{\theta}}_{y_{d,t}}\right\|_1\right)\right) \end{aligned} \quad (28)$$

For the same reason,

$$\begin{aligned} & \mathbb{E}_{q_{-y_{d,t}}}\left[\log\prod_{j=1}^{n_d}\left(p\left(u_{d,j}|\boldsymbol{\mu}_{y_{d,z_{d,j}},a_d}\right)\right)^{\mathbb{I}(t=z_{d,j})}\right] \\ & \quad = \sum_{j=1}^{n_d}\tilde{\boldsymbol{\phi}}_{d,j}(t)\sum_{u\in\mathcal{U}}\tilde{\boldsymbol{\upsilon}}_{d,j}(u)\left(\Psi\left(\tilde{\boldsymbol{\eta}}_{y_{d,t},a_d}(u)\right) - \Psi\left(\left\|\tilde{\boldsymbol{\eta}}_{y_{d,t},a_d}\right\|_1\right)\right) \\ & \mathbb{E}_{q_{-y_{d,t}}}\left[\log\prod_{j=1}^{n_d}\left(p\left(s_{d,j}|\boldsymbol{\sigma}_{y_{d,z_{d,j}},w_{d,j}}\right)\right)^{\mathbb{I}(t=z_{d,j})}\right] \\ & \quad = \sum_{j=1}^{n_d}\tilde{\boldsymbol{\phi}}_{d,j}(t)\sum_{s\in\mathcal{S}}\tilde{\boldsymbol{\rho}}_{d,j}(s)\left(\Psi\left(\tilde{\boldsymbol{\lambda}}_{y_{d,t},w_{d,j}}(s)\right) - \Psi\left(\left\|\tilde{\boldsymbol{\lambda}}_{y_{d,t},w_{d,j}}\right\|_1\right)\right) \end{aligned} \quad (29)$$

Plug (27), (28) and (29) back to (26),

$$\begin{aligned} \tilde{\boldsymbol{\xi}}_{d,t} \propto \exp\Bigg\{ & \Psi\left(\tilde{\boldsymbol{\alpha}}_{y_{d,t}}(1)\right) - \Psi\left(\left\|\tilde{\boldsymbol{\alpha}}_{y_{d,t}}\right\|_1\right) + \sum_{\ell=1}^{y_{d,t}-1}\Psi(\tilde{\boldsymbol{\alpha}}_\ell(2)) - \Psi(\|\tilde{\boldsymbol{\alpha}}_\ell\|_1) \\ & + \sum_{j=1}^{n_d}\tilde{\boldsymbol{\phi}}_{d,j}(t)\Bigg(\Psi\left(\tilde{\boldsymbol{\theta}}_{y_{d,t}}(w_{d,j})\right) - \Psi\left(\left\|\tilde{\boldsymbol{\theta}}_{y_{d,t}}\right\|_1\right) \\ & \qquad + \sum_{u\in\mathcal{U}}\tilde{\boldsymbol{\upsilon}}_{d,j}(u)\left(\Psi\left(\tilde{\boldsymbol{\eta}}_{y_{d,t},a_d}(u)\right) - \Psi\left(\left\|\tilde{\boldsymbol{\eta}}_{y_{d,t},a_d}\right\|_1\right)\right) \\ & \qquad + \sum_{s\in\mathcal{S}}\tilde{\boldsymbol{\rho}}_{d,j}(s)\left(\Psi\left(\tilde{\boldsymbol{\lambda}}_{y_{d,t},w_{d,j}}(s)\right) - \Psi\left(\left\|\tilde{\boldsymbol{\lambda}}_{y_{d,t},w_{d,j}}\right\|_1\right)\right)\Bigg)\Bigg\} \end{aligned} \quad (30)$$

The time complexity for this update is $O(KT^2D + KTN) = O(KTN)$ because $T \ll N$.

***Update of*** $\tilde{\boldsymbol{\phi}}$, the variational parameters for word topic assignment $z_{d,j}$, can be derived like (30) as the following with time complexity $O(KTN)$,

$$\begin{aligned} \tilde{\boldsymbol{\phi}}_{d,j} \propto \exp\Big\{ & \Psi\left(\tilde{\boldsymbol{\beta}}_{d,z_{d,j}}(1)\right) - \Psi\left(\left\|\tilde{\boldsymbol{\beta}}_{d,z_{d,j}}\right\|_1\right) + \textstyle\sum_{t=1}^{z_{d,j}-1}\Big(\Psi\left(\tilde{\boldsymbol{\beta}}_{d,t}(2)\right) - \\ & \Psi\left(\left\|\tilde{\boldsymbol{\beta}}_{d,t}\right\|_1\right)\Big) + \textstyle\sum_{k=1}^{K}\tilde{\boldsymbol{\xi}}_{d,z_{d,j}}(k)\Big(\Psi\left(\tilde{\boldsymbol{\theta}}_k(w_{d,j})\right) - \Psi\left(\left\|\tilde{\boldsymbol{\theta}}_k\right\|_1\right) + \\ & \textstyle\sum_{u\in\mathcal{U}}\tilde{\boldsymbol{\upsilon}}_{d,j}(u)\left(\Psi\left(\tilde{\boldsymbol{\eta}}_{k,a_d}(u)\right) - \Psi\left(\left\|\tilde{\boldsymbol{\eta}}_{k,a_d}\right\|_1\right)\right) + \\ & \textstyle\sum_{s\in\mathcal{S}}\tilde{\boldsymbol{\rho}}_{d,j}(s)\left(\Psi\left(\tilde{\boldsymbol{\lambda}}_{k,w_{d,j}}(s)\right) - \Psi\left(\left\|\tilde{\boldsymbol{\lambda}}_{k,w_{d,j}}\right\|_1\right)\right)\Big)\Big\} \end{aligned} \quad (31)$$

***Last, for update of*** $\tilde{\boldsymbol{\rho}}, \tilde{\boldsymbol{\upsilon}}$, the variational parameters for word sentiment assignments and preference assignments, one can use the techniques for previous inferences to arrive at,

$$\tilde{\boldsymbol{\rho}}_{d,j} \propto \exp\{\rho_{d,j}^{(1)} + \rho_{d,j}^{(2)}\}, \tilde{\boldsymbol{\upsilon}}_{d,j} \propto \exp\{\upsilon_{d,j}^{(1)} + \upsilon_{d,j}^{(2)}\}$$

$$\rho_{d,j}^{(1)} = \log r_d\, \mathbb{E}_{q_{-s_{d,j}}}[h_1] + \log(1-r_d)\, \mathbb{E}_{q_{-s_{d,j}}}[h_2] - \mathbb{E}_{q_{-s_{d,j}}}[\log B(h_1, h_2)]$$

$$\rho_{d,j}^{(2)} = \sum_{k=1}^{K}\sum_{t=1}^{T}\left(\Psi\left(\tilde{\boldsymbol{\lambda}}_{k,w_{d,j}}(s_{d,j})\right) - \Psi\left(\left\|\tilde{\boldsymbol{\lambda}}_{k,w_{d,j}}\right\|_1\right)\right)\tilde{\boldsymbol{\xi}}_{d,t}(k)\tilde{\boldsymbol{\Phi}}_{d,j}(t)$$
$$v_{d,j}^{(1)} = \log r_d\,\mathbb{E}_{q_{-u_{d,j}}}[h_1] + \log(1-r_d)\,\mathbb{E}_{q_{-u_{d,j}}}[h_2] - \mathbb{E}_{q_{-u_{d,j}}}[\log B(h_1,h_2)]$$
$$v_{d,j}^{(2)} = \sum_{k=1}^{K}\sum_{t=1}^{T}\left(\Psi\left(\tilde{\boldsymbol{\eta}}_{k,a_d}(u_{d,j})\right) - \Psi\left(\left\|\tilde{\boldsymbol{\eta}}_{k,a_d}\right\|_1\right)\right)\tilde{\boldsymbol{\xi}}_{d,t}(k)\tilde{\boldsymbol{\Phi}}_{d,j}(t) \quad (32)$$

where in $\rho_{d,j}^{(1)}$ and $v_{d,j}^{(1)}$, $h_1, h_2$ are the two dimensions of $h(\mathbf{s}_d, \mathbf{u}_d)$, and $B$ is the beta function. Recall the ratings are normalized in $[\epsilon, 1-\epsilon]$ where $\epsilon$ is a small positive decimal, so the logarithms in above equation are valid. It is usually very hard to exactly compute expectations in $\rho_{d,j}^{(1)}$ and $v_{d,j}^{(1)}$ due to the large space "$\mathbb{E}_{q_{-s_{d,j}}}$" needs to average over regardless of what $h$ we choose, let alone the beta function. By our best effort, using classic techniques like Stirling's approximation on beta function is also fruitless. Therefore, we resort to sampling w.r.t. density $q_{-s_{d,j}}$ and use the sample mean to estimate the expectations. Specially, for $h$ as in (2) that depends on $\mathbf{s}_d, \mathbf{u}_d$, the sampling is w.r.t. categoricals by $\tilde{\boldsymbol{\rho}}_{-d,j}$ and $\tilde{\boldsymbol{\upsilon}}_{-d,j}$. The sample size $m$ is a parameter of our variational model, and we recommend $m = 50$, which we find is a good balance between time and performance.

Our design of $h$ as in (2) involves mean and variance. An important caveat here is to prevent a zero or near-zero variance in (2) for short reviews with the highest/lowest rating, for which the $\boldsymbol{\rho}_d$ and $\boldsymbol{\eta}_d$ of that short review are very likely to place almost all their densities on extreme categories so that the corresponding $\mathbf{r}_d$ is almost constant. We thus should cap $h_1 + h_2 \le -\log\epsilon$. Another note is for switching the values of $s_{d,j}$ or $u_{d,j}$ to evaluate the categoricals $\tilde{\boldsymbol{\rho}}_{d,j}$ and $\tilde{\boldsymbol{\upsilon}}_{d,j}$. There is no need to re-compute the mean an variance every time; they can be updated in $O(1)$ time with formula in [16]. The time complexity for update of (32) is $O(m(|\mathcal{S}| + |\mathcal{U}|)N + KTN)$.

***Initialization***. Variational parameters can be initialized as their corresponding default parameters of the original model. For example, $\tilde{\boldsymbol{\alpha}}_k$ are all initialized to $(1,\alpha)$, $\tilde{\boldsymbol{\theta}}_k$ are all initialized to $\boldsymbol{\theta}$, $\tilde{\boldsymbol{\beta}}_{d,t}$ are all initialized to $(1,\beta)$, etc. For parameters like $\tilde{\boldsymbol{\xi}}, \tilde{\boldsymbol{\Phi}}$, they can be initialized using (30) (31). simplified by ruling out constant terms. More explanations can be found in [9].

*C. Stochastic Inference*

Even though variational inference developed in previous section is shown to have linear complexity in each iteration, it is still inefficient for very large corpus. The problem is that we need to update local variational variables for every document before we update the global variables. This is especially waste of effort at the beginning when global variables are not yet meaningful. Stochastic variational inference (SVI) improves the time use by digesting the documents individually. It has been shown in [9] that SVI achieves better performance given the same amount of time on large datasets.

We apply SVI to TSPRA to have a stochastic variational model *svTSPRA*, It can be summarized as the following after initializing all variables to $(\cdot)^{(0)}$. The experiment in section IV.A shows svTSPRA converges as fast as vTSPRA on much larger training sets.

- Repeat the following for $t = 1,2,3,\ldots$ until convergence
  - Randomly sample one document $d_t$, repeatedly compute the local variational parameters $\tilde{\boldsymbol{\xi}}^{(t)}, \tilde{\boldsymbol{\beta}}^{(t)}, \tilde{\boldsymbol{\Phi}}^{(t)}, \tilde{\boldsymbol{\rho}}^{(t)}, \tilde{\boldsymbol{\upsilon}}^{(t)}$ until convergence, using $\tilde{\boldsymbol{\alpha}}^{(t-1)}, \tilde{\boldsymbol{\theta}}^{(t-1)}, \tilde{\boldsymbol{\lambda}}^{(t-1)}, \tilde{\boldsymbol{\eta}}^{(t-1)}$.
    - Pretends document $d$ represents entire corpus $D$, compute intermediate global variables $\tilde{\boldsymbol{\alpha}}^*, \tilde{\boldsymbol{\theta}}^*, \tilde{\boldsymbol{\lambda}}^*, \tilde{\boldsymbol{\eta}}^*$ once, e.g. $\tilde{\boldsymbol{\theta}}_k^* = \boldsymbol{\theta} + D\sum_{t=1}^{T}\tilde{\boldsymbol{\xi}}_{d,t}(k)\sum_{j=1}^{n_{d_t}}\tilde{\boldsymbol{\Phi}}_{d,j}(t)\mathbf{e}_{w_{d,j}}$.
    - Merge $\tilde{\boldsymbol{\alpha}}^*, \tilde{\boldsymbol{\theta}}^*, \tilde{\boldsymbol{\lambda}}^*, \tilde{\boldsymbol{\eta}}^*$ into $\tilde{\boldsymbol{\alpha}}^{(t)}, \tilde{\boldsymbol{\theta}}^{(t)}, \tilde{\boldsymbol{\lambda}}^{(t)}, \tilde{\boldsymbol{\eta}}^{(t)}$ with some forgetting rate $r_t$, the weight on the intermediate values, to get $\tilde{\boldsymbol{\alpha}}^{(t+1)}, \tilde{\boldsymbol{\theta}}^{(t+1)}, \tilde{\boldsymbol{\lambda}}^{(t+1)}, \tilde{\boldsymbol{\eta}}^{(t+1)}$, e.g. $\tilde{\boldsymbol{\alpha}}^{(t+1)} = (1 - r_t)\tilde{\boldsymbol{\alpha}}^{(t)} + r_t\tilde{\boldsymbol{\alpha}}^*$. The experiments in [9] suggests we can intuitively set $r_t = \frac{1}{t+1}$.

Sometimes we might prefer analyzing the review data on the fly rather than waiting for a large corpus to be completely collected. We propose that SVI can be tuned for that purpose, and we present the following algorithm as *ovTSPRA*. We first let vTSPRA learn a base corpus of reasonable size $D_0$, and at time $t$ a new set of reviews $D_t$ arrive, where $D_0$ should be much larger than $D_t$. $D_0, D_t$ are used to denote both the corpus and their sizes in our discussion. The global variational parameters are $\tilde{\boldsymbol{\alpha}}^{(t)}, \tilde{\boldsymbol{\theta}}^{(t)}, \tilde{\boldsymbol{\lambda}}^{(t)}, \tilde{\boldsymbol{\eta}}^{(t)}$ at the end of time $t$. The process for time $t+1$ can be described Cas the following,

- Repeat the following until convergence
  - Compute the local variational parameters $\tilde{\boldsymbol{\xi}}^{(t+1)}, \tilde{\boldsymbol{\beta}}^{(t+1)}, \tilde{\boldsymbol{\Phi}}^{(t+1)}, \tilde{\boldsymbol{\rho}}^{(t+1)}, \tilde{\boldsymbol{\upsilon}}^{(t+1)}$repeatedly until convergence, using $\tilde{\boldsymbol{\alpha}}^{(t)}, \tilde{\boldsymbol{\theta}}^{(t)}, \tilde{\boldsymbol{\lambda}}^{(t)}, \tilde{\boldsymbol{\eta}}^{(t)}$.
    - Pretends $D_{t+1}$ has size $D_0$, e.g. each review in $D_{t+1}$ is treated as $\frac{D_0}{D_{t+1}}$ reviews, and then compute the intermediate $\tilde{\boldsymbol{\alpha}}^*, \tilde{\boldsymbol{\theta}}^*, \tilde{\boldsymbol{\lambda}}^*, \tilde{\boldsymbol{\eta}}^*$ for $D_t$ until convergence and maximum $D_{t+1}$ rounds iterations.
    - Merge the intermediate $\tilde{\boldsymbol{\alpha}}, \tilde{\boldsymbol{\theta}}, \tilde{\boldsymbol{\lambda}}, \tilde{\boldsymbol{\eta}}$ for $D_{t+1}$ into $\tilde{\boldsymbol{\alpha}}^{(t)}, \tilde{\boldsymbol{\theta}}^{(t)}, \tilde{\boldsymbol{\lambda}}^{(t)}, \tilde{\boldsymbol{\eta}}^{(t)}$ with some forgetting rate $r_t$ to get $\tilde{\boldsymbol{\alpha}}^{(t+1)}, \tilde{\boldsymbol{\theta}}^{(t+1)}, \tilde{\boldsymbol{\lambda}}^{(t+1)}, \tilde{\boldsymbol{\eta}}^{(t+1)}$. We can intuitively let $r_t = \frac{D_{t+1}}{D_0}$ for the general case, or higher $r_t$ if new data are considered more informative.

ovTSPRA is lesser in its predictive power if given an randomly chosen test set, but it has several advantages. 1) Its parameter evaluation better reflects the recent trends and our experiments in IV.A show it exhibits better or competitive performance when used to predict recent reviews. 2) It is an online algorithm and can be used to monitor the oscillation in sentiment and preference or any other time-series analysis, can capture unusual fluctuations in a timely manner. Some meaningful results are demoed in section IV.C. 3) If we are only interested in recent trends and the near future, which is a common application, ovTSPRA is the most appropriate one with clearly lowest computation cost.

### D. Prediction

For prediction, global variables $\mathbf{b}, \boldsymbol{\varphi}, \boldsymbol{\sigma}, \boldsymbol{\mu}$ are assumed known because they can be trained by the method in the previous section. On the contrary, review ratings become unobservable and their values are to be predicted. The plate diagram of prediction is depicted in Figure 2. By this change, variational parameters $\tilde{\boldsymbol{\alpha}}, \tilde{\boldsymbol{\theta}}, \tilde{\boldsymbol{\lambda}}$ are no longer needed, while $D \times 2$ new variational parameters $\tilde{\mathbf{r}} = (\tilde{\mathbf{r}}_1, \dots, \tilde{\mathbf{r}}_D)$ are introduced for review ratings and we assume $r_d \sim \text{beta}(\tilde{\mathbf{r}}_d)$. The updates for $\tilde{\boldsymbol{\beta}}$ stays the same as (25), while $\tilde{\boldsymbol{\xi}}, \tilde{\boldsymbol{\phi}}$ and $\rho_{d,j}^{(2)}, \upsilon_{d,j}^{(2)}$ in (32) maintain their structures with all digamma functions $\psi(\cdot) - \psi(\|\cdot\|)$ involving $\tilde{\boldsymbol{\alpha}}, \tilde{\boldsymbol{\theta}}, \tilde{\boldsymbol{\lambda}}$ replaced by $\frac{(\cdot)}{\|\cdot\|_1}$ as the following,

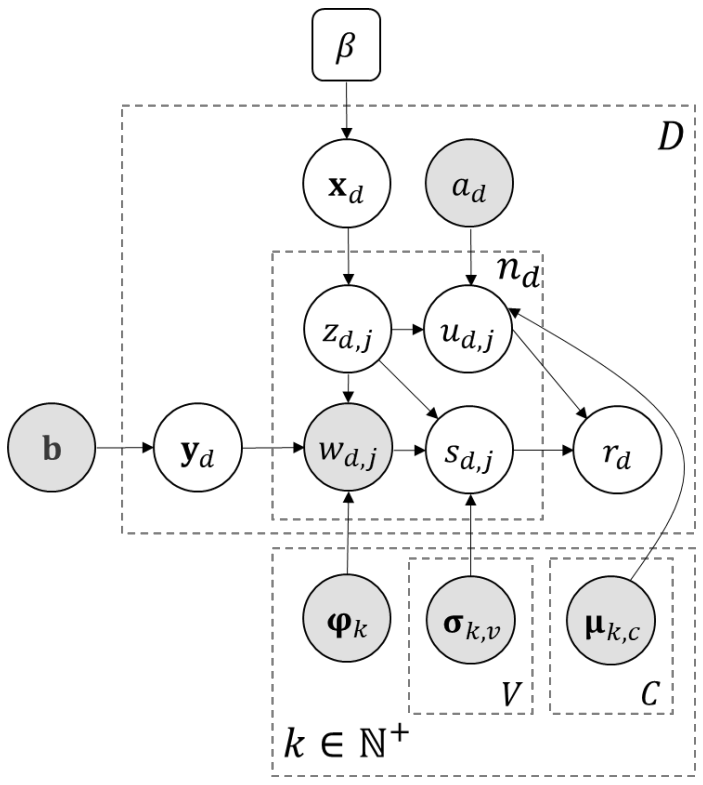

Figure 2 The plate diagram for vTSPRA prediction model.

$$
\begin{aligned}
\tilde{\boldsymbol{\xi}}_{d,t} \propto \exp & \left\{ \frac{\tilde{\boldsymbol{\alpha}}_{y_{d,t}}(1)}{\|\tilde{\boldsymbol{\alpha}}_{y_{d,t}}\|_1} + \sum_{\Bbbk=1}^{y_{d,t}-1} \frac{\tilde{\boldsymbol{\alpha}}_{\Bbbk}(2)}{\|\tilde{\boldsymbol{\alpha}}_{\Bbbk}\|_1} \right. \\
& \left. + \sum_{j=1}^{n_d} \tilde{\boldsymbol{\phi}}_{d,j}(t) \left( \frac{\tilde{\boldsymbol{\theta}}_{y_{d,t}}(w_{d,j})}{\|\tilde{\boldsymbol{\theta}}_{y_{d,t}}\|_1} + \sum_{u \in \mathcal{U}} \frac{\tilde{\boldsymbol{\upsilon}}_{d,j}(u) \tilde{\boldsymbol{\eta}}_{y_{d,t},a_d}(u)}{\|\tilde{\boldsymbol{\eta}}_{y_{d,t},a_d}\|_1} + \sum_{s \in \mathcal{S}} \frac{\tilde{\boldsymbol{\rho}}_{d,j}(s) \tilde{\boldsymbol{\lambda}}_{y_{d,t},w_{d,j}}(s)}{\|\tilde{\boldsymbol{\lambda}}_{y_{d,t},w_{d,j}}\|_1} \right) \right\} \\
\tilde{\boldsymbol{\phi}}_{d,j} \propto \exp & \left\{ \Psi\left(\tilde{\boldsymbol{\beta}}_{d,z_{d,j}}(1)\right) - \Psi\left(\|\tilde{\boldsymbol{\beta}}_{d,z_{d,j}}\|_1\right) + \sum_{t=1}^{z_{d,j}-1} \left( \Psi\left(\tilde{\boldsymbol{\beta}}_{d,t}(2)\right) - \Psi\left(\|\tilde{\boldsymbol{\beta}}_{d,t}\|_1\right) \right) \right. \\
& \left. + \sum_{k=1}^{K} \tilde{\boldsymbol{\xi}}_{d,z_{d,j}}(k) \left( \frac{\tilde{\boldsymbol{\theta}}_k(w_{d,j})}{\|\tilde{\boldsymbol{\theta}}_k\|_1} + \sum_{u \in \mathcal{U}} \frac{\tilde{\boldsymbol{\upsilon}}_{d,j}(u) \tilde{\boldsymbol{\eta}}_{k,a_d}(u)}{\tilde{\boldsymbol{\eta}}_{k,a_d}} + \sum_{s \in \mathcal{S}} \frac{\tilde{\boldsymbol{\rho}}_{d,j}(s) \tilde{\boldsymbol{\lambda}}_{k,w_{d,j}}(s)}{\|\tilde{\boldsymbol{\lambda}}_{k,w_{d,j}}\|_1} \right) \right\} \\
\rho_{d,j}^{(2)} = & \sum_{k=1}^{K} \sum_{t=1}^{T} \frac{\tilde{\boldsymbol{\lambda}}_{k,w_{d,j}}(s_{d,j}) \tilde{\boldsymbol{\xi}}_{d,t}(k) \tilde{\boldsymbol{\phi}}_{d,j}(t)}{\|\tilde{\boldsymbol{\lambda}}_{k,w_{d,j}}\|_1}, \upsilon_{d,j}^{(2)} = \sum_{k=1}^{K} \sum_{t=1}^{T} \frac{\tilde{\boldsymbol{\eta}}_{k,a_d}(u_{d,j}) \tilde{\boldsymbol{\xi}}_{d,t}(k) \tilde{\boldsymbol{\phi}}_{d,j}(t)}{\|\tilde{\boldsymbol{\eta}}_{k,a_d}\|_1}
\end{aligned}
\tag{33}
$$

The update $\rho_{d,j}^{(1)}, \upsilon_{d,j}^{(1)}$ in (32) need to consider $\tilde{\mathbf{r}}_d$, evaluated by the same sampling as described in previous section.

$$
\begin{aligned}
\rho_{d,j}^{(1)} = & \left(\Psi(\tilde{\mathbf{r}}_d(1)) - \Psi(\|\tilde{\mathbf{r}}_d\|_1)\right) \mathbb{E}_{q-s_{d,j}}[h_1] \\
& + \left(\Psi(\tilde{\mathbf{r}}_d(2)) - \Psi(\|\tilde{\mathbf{r}}_d\|_1)\right) \mathbb{E}_{q-s_{d,j}}[h_2] - \mathbb{E}_{q-s_{d,j}}[\log B(h_1, h_2)] \\
\upsilon_{d,j}^{(1)} = & \left(\Psi(\tilde{\mathbf{r}}_d(1)) - \Psi(\|\tilde{\mathbf{r}}_d\|_1)\right) \mathbb{E}_{q-u_{d,j}}[h_1] \\
& + \left(\Psi(\tilde{\mathbf{r}}_d(2)) - \Psi(\|\tilde{\mathbf{r}}_d\|_1)\right) \mathbb{E}_{q-u_{d,j}}[h_2] - \mathbb{E}_{q-u_{d,j}}[\log B(h_1, h_2)]
\end{aligned}
\tag{34}
$$

The update for $\tilde{\mathbf{r}}_d$ can be solved as the following, evaluated by the same samples.

$$
\tilde{\mathbf{r}}_d = \log(r_d) \mathbb{E}_{q_{-r_d}}[h_1 - 1] + \log(1 - r_d) \mathbb{E}_{q_{-r_d}}[h_2 - 1] \tag{35}
$$

Finally, the prediction $\hat{r}_d$ for rating $r_d$ is $\hat{r}_d = \tilde{\mathbf{r}}_d(1)$. The prediction is mathematically elegant without setting in many heuristics like [3].

## IV. Experiment & Evaluation

We conduct our experiments with matlab on a i7 3.0GHz computer. We use eight public Amazon data sets [17] "Book", "Movie", "Music", "Cellphone", "Clothing", "Pet", "Automobiles", "Arts" to evaluate our proposed models. Each data set is a collection of reviews with each review associated with an author ID, a product ID and a rating. The rating is a 1-5 scale to indicate the customer's satisfactory level for the purchase. We only consider authors who post more than 1 reviews. We use Stanford NLP toolkit [18] to pre-process the review texts to 1) convert all words to their lemma from, so that like a none and its plural, a verb and its past tense, etc., are treated as the same word; 2) symbols, cardinal numbers, determiners, pronoun, most conjunctions, a few stop words, and any lemma of less than 3 letters or occurrences less than 10 times in the whole corpus are removed. Some descriptive dataset description after pre-processing are shown in Table 2. We can see the first three are large datasets and the remaining are (relatively) small datasets.

| | $D$ | $N$ | $C$ | $V$ | | $D$ | $N$ | $C$ | $V$ |
|---|---|---|---|---|---|---|---|---|---|
| Book | 343 | 36821 | 65 | 16 | Cloth | 2.2 | 96 | .94 | 1.1 |
| Movie | 180 | 19635 | 32 | 14 | Pet | 6.4 | 382 | 2.3 | 1.8 |
| Music | 182 | 18232 | 30 | 11 | Auto | 4.9 | 248 | 1.8 | 1.4 |
| Phone | 1.2 | 102 | .52 | .78 | Arts | .28 | 15 | .12 | .82 |

Table 2 Dataset descriptive statistics. All numbers in $\times \mathbf{10^4}$ scale. $\boldsymbol{D}$: the number of reviews; $\boldsymbol{N}$: the total number of lemmatized words; $\boldsymbol{C}$: the number of authors; $\boldsymbol{V}$: the vocabulary size.

We notice the word sentiment distribution evaluated by TSPRA in [3] is ***positively skewed***, because the Amazon datasets typically contain more positive reviews (4/5-star) than negative and neutral reviews (1/2/3-star). This will be worse for large datasets, as 80% to 90% of their reviews are positive, and their size can skew the distribution even more. Our solution is to calculate the ratio of positive reviews over negative/neutral reviews, and "replicate" each non-negative review by this ratio in the corpus. This "balance" is not just technically need, it is also reasonable from application's perspective, because typically negative and neutral reviews are more concerned by online vendors. The "balanced" corpus help us see more clearly which topic has problem form the analysis, thus all following experiments are based on such "balanced" corpus.

### A. Performance Analysis

We first measure the prediction performance of our new models on above-mentioned eight datasets in comparison with TSPRA and FLAME in terms of prediction error. All models use their respective recommended initial parameters and are given 48 hours to run. For our model, $K = 100, T = 10, \alpha = \beta = 1 + 10^{-10}$ (with the small decimal to prevent infinity in digamma function for short reviews), uniform $\boldsymbol{\lambda}, \boldsymbol{\eta}$, and variational parameter initialization as at the end of section III.B. The normalized rating is restored to the original scale 1 to 5 and measure the error.

For every dataset, reviews are ordered by time, and the first 80% are used for training with the remaining for test. For three large datasets, we randomly draw about 300,000 of the reviews from training set to feed in TSPRA and FLAME, and about 600,000 to feed in svTSPRA, which are reasonable size for them to reach good convergence; svTSPRA runs over entire training set.

We also experiment on ovTSPRA in terms of prediction error. Recall that it should predict the near future. For three large review set, we set the base size as 20,000 reviews at the beginning of timeline and use 2,000 as step size, and for each step we predict the next 2,000 reviews; as to other three small review sets, we use one tenth of above quantities. The prediction errors are averaged as the final results. We have following conclusions,

1) We confirm that TSPRA outperforms FLAME on all the tested large and small data sets.
2) On the large datasets, svTSPRA clearly outperforms other models, because of its capability of learning more training data in the same time to achieve better evaluated parameters and miss much less words and users in the test set. On small data sets, since Gibbs sampling in TSPRA is able to process all training data and reach steady state in time, then in general it runs better than the variational models.
3) Meanwhile, we observe that ovTSPRA exhibits smaller or competitive error against TSPRA in its near-future prediction.

| | ovTSPRA | svTSPRA | vTSPRA | TSPRA | FLAME |
|---|---|---|---|---|---|
| Book | **0.694**(2) | **0.680**(1) | 0.796 | 0.983 | 1.133 |
| Movie | **0.633** | **0.671** | 0.759 | 0.875 | 0.924 |
| Music | **0.739** | **0.762** | 0.872 | 0.982 | 1.221 |
| Phone | **0.818** | 0.984 | 0.998 | 0.935 | 1.108 |
| Cloth | *0.414*(2) | **0.307** | 0.337 | 0.310 | 0.528 |
| Pet | **0.663** | 0.829 | 0.803 | 0.693 | 0.847 |
| Auto | *0.634* | 0.696 | 0.653 | 0.578 | 0.631 |
| Arts | *0.648* | 0.701 | 0.719 | 0.600 | 0.642 |

Table 3 Prediction performance evaluation results. It measures prediction errors of each model on the test set w.r.t. the original rating scale 1-5, except for ovTSPRA . (1) svTSPRA runs better than other models on large datasets. (2) ovTSPRA in its own near-future prediction gives better or competitive error against TSPRA and svTSPRA.

## B. Convergence Analysis

The convergence of svTSPAR, vTSPAR and TSPRA on datasets Book, Movie, Music in terms of rating regression error on their training sets are given in Figure 3 (a), (b), (c). We start the measurement of variational models after 5hrs, and TSPRA after its 500 burn-in iterations (about 25hrs). It shows svTSPRA converges as fast as vTSPRA on its much larger training sets, to a competitive or better regression error.

The time usage of one iteration of vTSPRA is briefly experimented on complete training sets to confirm a linear time complexity as analyzed in III.B. We plot them in Figure 3 (d) with a log10-log10 scale, x-axis being the number of words, y-axis being the runtime in seconds.

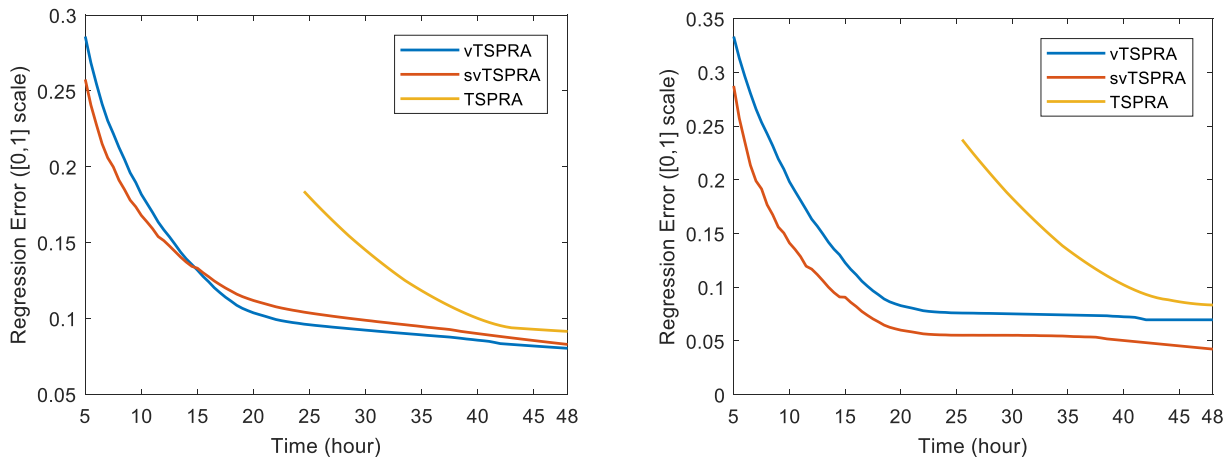

(a) (b)

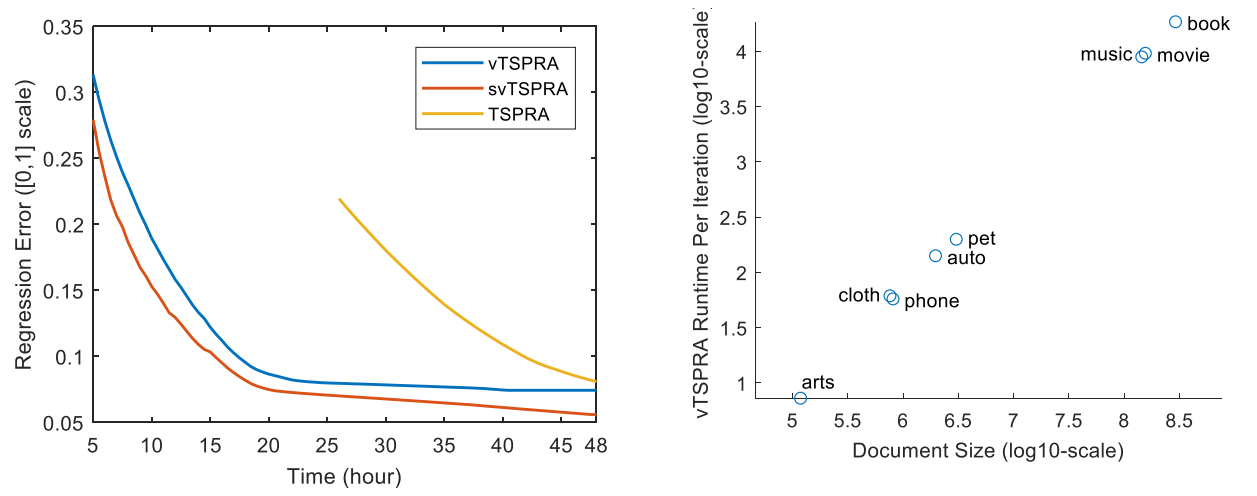

(c) (d)

Figure 3 (a), (b), (c) convergence of variational TSPRA models and TSPRA itself on their training sets. (d) one iteration time of vTSPRA on the whole training set.

## C. Sentiment Annotation

Our model can be used to automatically generate word sentiment annotation, a distinct feature from previous models like [1] or [2]. We empirically compare our sentiment annotation with the latest SenticNet4 [14], a public word sentiment annotation set whose sentiments are evaluated by linguistic methods. Our word sentiments are aggregated across all datasets, and we only consider words with more than 20 occurrences and those also in SenticNet4. The sentiment distribution is plotted in Figure 4, and we observe clear difference that our distribution is more Gaussian-like, with a slightly positive mean 0.1.

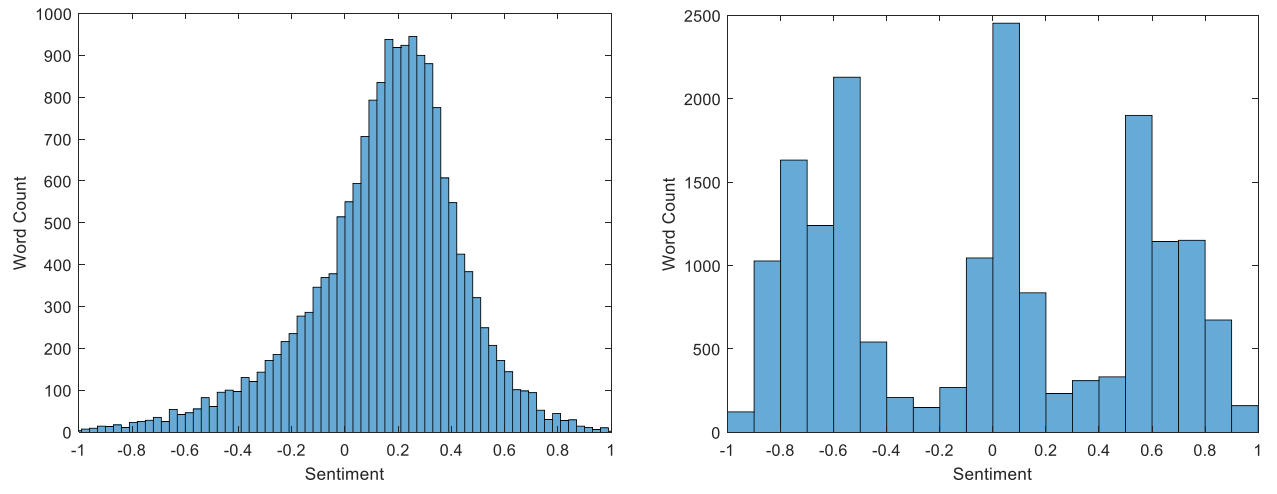

(a) (b)

Figure 4 Word sentiment distribution by svTSPRA and SenticNet 4.

Our annotation is more appropriate in the review context. SenticNet tend to give more extreme ratings for intuitively neutral words in review context, especially product names and features. Some examples, first number being our annotation and the second being SenticNet: “ingenuousness” (0.32/-0.59), “numeration” (-0.12/0.74), “translucence” (0.16/-0.66), “sudokus” (0/0.59), “gangway” (0.11/0.83), “gherkin” (-0.13/-0.66), etc.

For quantitative comparison, we plug the sentiment values of SenticNet 4 into the prediction of svTSPRA to replace the word sentiment values derived from learning, and run over three large datasets again. There is no surprising the error significantly increases, as shown in Table 4.

| | Book | Movie | Music |
|---|---|---|---|
| SenticNet 4 | 0.947 (+39%) | 1.113 (+66%) | 1.154 (+51%) |
| svTSPRA | 0.680 | 0.671 | 0.762 |

Table 4 The comparison of using sentiment values from SenticNet 4 and learned values from svTSPRA in prediction. On average

## D. Sentiment & Preference Trends

Another distinct application of our model is ovTSPRA can monitor the change in sentiment and preference over time. We now present several topic-level sentiments and user preference (see equation (17)) fluctuations captured by ovTSPRA and provide our interpretations and possible explanations.

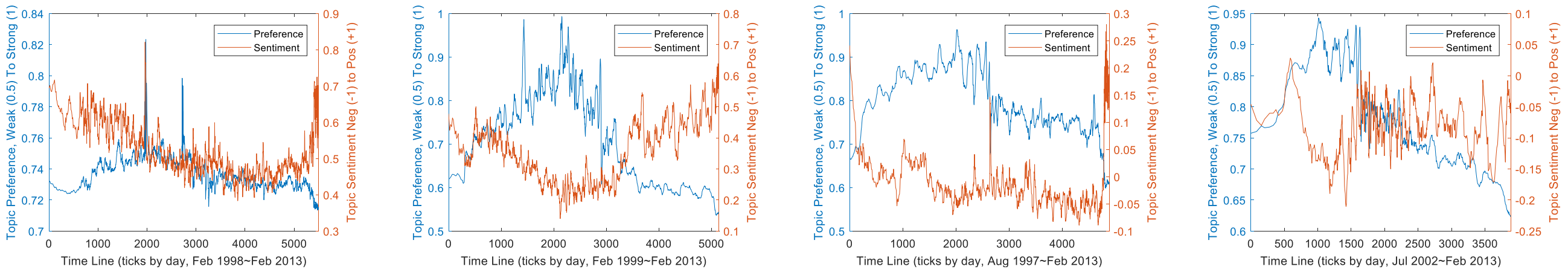

(a) Book, Harry Potter (b) Music, hip-hop (c) Movie, romantic (d) Cellphone, headset

Figure 5 Examples of some topic-level sentiment and preference trends analyzed by ovTSPRA.

1) One interesting topic identified by our model in the Book dataset is mainly about Harry Potter series, represented by words “harry”, “rowling”, “muggle”, “hogwarts”, “voldemort”, “snape”, “wizard”, “magic”, “magician”, “wand”, “quidditch”, “half-blood”, etc. This book series maintains a relatively high user preference and sentiment through the years, although the sentiment has a downward trend. The two peaks in the curves seem to correspond to the publication of the series’ fifth and sixth book.
2) The hip-hop topic in music is represented by words “hip-hop”, “rap”, “rapper”, “beat”, “r&b”, “emcee”, “tribe”, “rhyme”, “urban”, “gangsta”, “clan”, etc. and some popular singers’ names. This topic draws our attention because of its contradicting trends of sentiment and preference. It is possible that the hip-hop fans tend to be somewhat negative when they concern this topic and post reviews online.
3) Romantic movies are represented by words “rom-com”, “romance”, “wedding”, “comedy”, “relationship”, “love”, “divorce”, “triangle”, “chemistry”. It enjoys a rise of user preference in early 2000s, with its peak at late 2002 to early 2003. However, its overall sentiment is almost always around neutral.
4) The cellphone headset topic is represented by word “headset”, “ear”, “sound”, “bluetooth”, “quality”, “call”, “volume”, “noise”, etc. From 2005 to 2006, users have a growing concern of headset, probably because of the emergence of smart phone. Nonetheless, the growth is associated with a drop in sentiment, reflecting that people are not satisfied with headset quality in the early years. Later, user preference decreases, probably because users start to concern more about other aspects of cellphone, but in contrast the topic sentiment improves.

For one thing, above examples demo an interesting application of our model, for the other thing it gives strong visual evidence that user preference is for the most part not correlated with sentiment and is better treated as an independent factor in review models, as we can see the oscillation of sentiment and preference in can be in the same direction and the opposite direction.

## V. Conclusion

This paper proposes and develop several joint topic-sentiment-preference analysis models vTSPRA, svTSPRA, ovTSPRA for online reviews under the variational inference framework. The original TSPRA model is completely reconstructed with a new generative process with improved regression design. We then present linear-time coordinate updates of variational parameter and adapt them to stochastic algorithms, enabling the models to process large review sets of millions of reviews and achieve better performance by learning from complete training set rather than a small sample. The online algorithm ovTSPRA also finds its application in monitoring sentiment and preference change, and several interesting examples have been demoed.